\documentstyle[preprint,pra,aps]{revtex}
\begin{document}

\draft

\title{Quantum Cryptography with Coherent States}

\author{B. Huttner and N. Imoto}
\address{NTT Basic Research Laboratories\\
3-1 Morinosato Wakamiya, Atsugi, Kanagawa 243-01, Japan}

\author{N. Gisin}
\address{Group of Applied Physics, University of Geneva\\
CH-1211 Geneva 4, Switzerland}

\author{T. Mor}
\address{Department of Physics, Technion - Israel Institute
of Technology\\ 32000 Haifa, Israel}

\date{ }

\maketitle

\begin{abstract}
The safety of a quantum key distribution system relies on the
fact that any eavesdropping attempt on the quantum channel
creates errors in the transmission. For a given error
rate, the amount of information that may have leaked to
the eavesdropper depends on both the particular system
and the eavesdropping strategy.
In this work, we discuss quantum cryptographic protocols
based on the transmission of weak coherent states and
present a new system,
based on a symbiosis of two existing ones, and for which
the information available to the eavesdropper is significantly
reduced.
This system is therefore safer than the two previous ones.
We also suggest a possible
experimental implementation.
\end{abstract}


\section{Introduction}
\label{intro}

The only known method to exchange secret information through a
communication channel in a proven secure way,
is to use the so-called one-time pad (for a good review of both classical and
quantum cryptography, see~\cite{bra88}).
In this technique, the data,
which is represented by a string of bits, is combined with a
random string of bits of equal length called the key, and is
then sent through the communication channel.
The randomness of the key ensures that the encoded message is also completely
random and as such totally unintelligible to a potential
eavesdropper. The safety of the transmission is thus
entirely dependent on the safety of the key, which has to
be secret
and shared just by both legitimate users. Moreover, safety can
be guaranteed only if the key is used once, and then
discarded.
The problem is therefore how to distribute the random key
between users in a secure way. Classically, the only
possibility is either through personal meetings, or through a
trusted courier, which
makes the technique rather expensive, and not practical for many
applications.
Therefore, most practical cryptographic systems nowadays rely
on different principles~\cite{bra88}. However, these cannot
really {\em guarantee} the safety of the transmission,
but rely on a
weaker property of the system, namely that it is
{\em computationally} safe.
This means that the system can be broken in principle, but
that the computation time required to do so is too long to
pose a real threat. The main problem with this approach is that its safety
could be destroyed by technological progress (faster computers)
or mathematical advances (faster algorithms or future
theoretical progress in computation theory).
Another technique, whose safety does not rely on computing
abilities, and which was only recently
developed, is quantum cryptography (for an
introduction, see~\cite{review}).

In quantum cryptography, the two users, generally referred to as
Alice (the sender) and Bob (the receiver),
have two kinds of communication channels at their
disposal. One is a classical public channel, which can be
overheard by anybody, but cannot be modified; and the second
is a quantum channel, whose main characteristic is that any
attempt at eavesdropping will create errors in the transmission.
The quantum channel will be used to transmit the secret key,
and the classical public channel will be used to exchange
information and to send the encoded message.
In principle, this is sufficient to ensure the
safety of a transmission: Alice and Bob exchange a series of
bits over the quantum channel, and then use part of the
transmission to test for eavesdropping. If they find any
discrepancy between their strings, they can infer that an
eavesdropper, usually referred to as Eve, was
listening and that their transmission is not secret. If
they detect no errors, they can assume that the key is safe.
By testing a large proportion of their initial string, they can
attain any safety level they wish.
Unfortunately, quantum channels are very sensitive devices,
and due to the imperfections of the channels and of the
detectors, some errors will always be unavoidable. The problem
facing Alice and Bob is therefore, for a given error rate,
to estimate  the amount of information that may have leaked to Eve,
and decide on the safety of the transmission. This of
course depends on both the particular system used by Alice and
Bob, and on the eavesdropping strategy adopted by Eve. A safer
system is a system for which the amount of information that may have
leaked to Eve is lower. If the information leaked to Eve
is not too high, Alice and Bob can use classical information
processing techniques to reduce it to approximately zero,
at the expense of shortening their
strings~\cite{ben92,ben88,hut94}.

At present, there exist three different quantum cryptographic
systems. The first one relies
on the transmission of single photons randomly polarized
along four directions~\cite{ben84}. As single photons are
difficult to produce experimentally, a slight
modification of this system, using weak pulses instead of
single photons,  was the first one to be implemented in
practice~\cite{ben92,tow93,mul93}. The second system,
which is conceptually the simplest, uses only two
non-orthogonal quantum states~\cite{bena92}.
Its implementation relies on
weak coherent pulses, with a phase difference between
them~\cite{bena92,towa93}.
The third system  is based on the creation of pairs
of EPR correlated photons~\cite{eke91}.
One of its potential advantages
is that the correlations are between single photons and not
weak pulses, which can be a great advantage, as we shall
emphasize later. However, creation and transmission over
long distances of EPR correlated pairs
is technologically more difficult, and it is not clear yet
whether this will prove practical~\cite{eke92}.
In this work, we focus on quantum cryptographic schemes
implemented with weak pulses of coherent light.
We compare the safety of the first and the
second of these quantum
cryptographic systems, and present a new system, which is a
symbiosis of both, and for which the safety can be
significantly increased.

In Section~\ref{4}, we analyze the first system, referred to
as 4-states system. In Section~\ref{2}, we turn to the second
one, named 2-states system, and present a new implementation.
We introduce our new 4+2 system in Section~\ref{4+2}, and show
that it is more sensitive to eavesdropping than the two
previous ones. In Section~\ref{lossy}, we show the dangers
associated with a lossy transmission line, and conclude in
Section~\ref{conclusion}.

\section{4-states protocol}
\label{4}

\subsection{Principle of 4-states systems}
This protocol was developed by Bennett et
al.~\cite{ben84}.
 The sender, Alice, chooses at random one out
of four states, e.g. for polarized photons:
\mbox{$\updownarrow$}, \mbox{$\leftrightarrow$},
\mbox{$\nearrow\kern-1.02em\swarrow$} or
\mbox{$\searrow\kern-1.02em\nwarrow$},
and sends it to the receiver, Bob. The two states
\mbox{$\updownarrow$} and \mbox{$\nearrow\kern-1.02em\swarrow$}
stand for bit value `0', while the other two,
\mbox{$\leftrightarrow$} and
\mbox{$\searrow\kern-1.02em\nwarrow$}, stand for `1'.
Bob chooses, also at
random, a basis, {\mbox{$\oplus$} } or
\mbox{$\otimes$}, in which he measures the
polarization. When his basis corresponds to Alice's, his bit
should be perfectly correlated with  hers, whereas when
his basis is the conjugate, there is no correlation between
his result and Alice's original choice. By discussing over the
public channel, Alice and Bob agree
to discard all the instances where they did not use the same
basis (half of the total on average).
The result is what we call the sifted key, which should be
two perfectly correlated strings, but which may contain errors.
The two fundamental
properties of this protocol are:
\begin{itemize}
\item[(i)] the choice of basis is completely hidden from the other
protagonist (the two bases correspond to the same density
matrix), as well as from any mischievous eavesdropper,
Eve;
\item[(ii)] when Alice and Bob use different bases, there is no
correlation between their bits.
\end{itemize}
The first one, (i), ensures that,
as the eavesdropper Eve cannot know which basis was used,
she will unavoidably  introduce errors. The second one, (ii),
is not really necessary, but is preferable, as it reduces
the information available to Eve to a minimum~\cite{pho93}
(25 \% for each photon on which she eavesdropped).
There  have been various works
analyzing eavesdropping strategies, calculating the information
available to Eve as a function of the error rate and developing
information processing techniques to reduce it to any required
level~\cite{ben92,ben88,hut94,bar94}.

\subsection{Implementation with weak pulses}
One technical difficulty with this scheme is that in principle
it should be implemented by means of
single photons~\cite{ben84}. As these
are difficult to generate experimentally, existing
schemes rely on weak pulses of coherent
or thermal light, with
much less than one photon per pulse on
average~\cite{ben92,tow93,mul93}. This ensures
that the probability of having two or more photons in a pulse
remains very small. This strategy reduces the transmission rate
(recent experiments use about one tenth of a photon per pulse),
while providing no advantage to an honest participant. More
precisely, if Alice and Bob use coherent pulses
$| \alpha \rangle$, the transmission rate $t^{(4)}$ is given by:
\begin{equation}
\label{t4}
t^{(4)} \equiv  \frac{1}{2}
\left( 1- \left| \langle \alpha |0 \rangle \right| ^2  \right)
  =  \frac{1}{2} \left( 1 - e^{-|\alpha|^2} \right) \; ,
\end{equation}
where the factor $1/2$ comes from the fact that half of
the transmissions had to be discarded (when Alice and Bob
used different bases).  In fact,
even for these weak pulses, the probability of having two or
more photons per pulse may not always be neglected
(for the above
pulses, one out of twenty non-zero pulses will have two
photons). We shall show in Section~\ref{lossy} how to take
this into
account.

\subsection{Safety of 4-states systems}
In order to get quantitative results, we shall assume that Eve
uses the intercept/resend strategy: she intercepts the pulses,
attempts to gain as much information as possible,
and sends to Bob a new pulse, prepared according to the
information she obtained. Moreover, we shall assume that  she eavesdrops in
the bases used by Alice and Bob, {\mbox{$\oplus$} } or
\mbox{$\otimes$}.
This is the intercept/resend strategy which
provides her with the most information on the sifted
key~\cite{hut94}. However, it is not yet known whether
this is the optimal
strategy.
It is easy to see that, when Eve eavesdrops on a fraction $\eta$
of the transmissions, the error rate created is
$\eta/4$ (when Eve uses the correct basis, she does not
introduce any error, while she creates a 50\% error rate when
she uses the wrong basis), and that the information she
obtained is $\eta/2$ (she has total information when
she used the correct basis, and none when she used the wrong
one). Moreover, the scheme is completely symmetric, so that Eve
shares the  same information with Alice and with Bob. Therefore,
we can write the mutual information shared by Alice and Eve and
shared by Eve and Bob as a function of the error rate $Q$:
\begin{equation}
\label{i4}
I_{AE}^{(4)} (Q) = I_{EB}^{(4)} (Q) = 2 Q \; .
\end{equation}
In this system, the intensity of the weak pulses, or
equivalently the transmission rate $t^{(4)}$ defined
in~(\ref{t4}) has no influence.
In the following, we will compare~(\ref{i4})
to the information
obtained by Eve for the other two systems.

\section{2-states protocol}
\label{2}

\subsection{Principle of 2-states systems}
In this protocol~\cite{bena92},  Alice chooses between only
two non-orthogonal
states, and sends one to Bob. As these are not orthogonal,
there is no way for Bob to decode them deterministically.
However, by means of a generalized measurement, also known as
{\em positive operator valued measure} or
POVM~\cite{per93,eke94}, he can
perform a test which will sometimes fail to give an answer,
and at all other times give the correct one. In essence, instead
of having a binary test (with results 0 or 1),
which will unavoidably create errors
when the two states are not orthogonal, Bob has a ternary
system, with possible results: 0, 1, or ? (with ? corresponding
to inconclusive results). For example, if Alice sends a 0,
Bob may get
either a 0 or an inconclusive result, but he will never get a 1.
We present a practical implementation of such a POVM in the
next Section (for a theoretical description
see~\cite{per93,eke94}). As with the previous system,
Alice and Bob use their public
channel to discard all the inconclusive results. They should
now have two perfectly correlated strings, except for possible
errors.
The safety of the protocol against
eavesdropping is ensured by the fact that Eve cannot get
deterministic results either. She may attempt to get as much
information as possible by projecting the states onto an
orthogonal basis. This is discussed in Section~\ref{safety2},
where we show that this will unavoidably create errors.
Another possibility for Eve is to mimic Bob's measurement, and
obtain deterministic results on a fraction of the bits.
However, in this case, she will have to guess Alice's choice
on the remaining bits, and this will provide her with less
information than the previous strategy.
Let us emphasize that, as we want Eve's guesses to result
in errors on Bob's side, this scheme should not be implemented
with weak pulses only.
In this case, Eve could simply
intercept the transmission, resend a pulse to Bob only
when she managed to obtain the bit sent by Alice
(i.e. 0 or 1), and send nothing when she obtained
an inconclusive result. The
signature of eavesdropping would then just be a
reduction in the
transmission rate. In the original proposal, this is
overcome by using phase encoding of a weak pulse
(the two states are $| \pm \alpha \rangle$, with overlap
$e^{-2|\alpha|^2}$), which is
sent together with a strong pulse, used also as a phase
reference~\cite{bena92,towa93}.
When Eve fails to obtain information, she still has
to send the strong pulse, which  will create errors in the
reception. Ekert et al.~\cite{eke94} recently analyzed
various eavesdropping
strategies related to this system. Here, we shall restrict
ourselves to the intercept/resend strategy and compare the
safety of various schemes.

\subsection{New implementation}
Let us now suggest a new implementation of the
2-states system, which will be later extended to our
new 4 + 2 protocol. We shall call it {\em parallel reference}
implementation, to distinguish it from the
{\em sequential reference}
implementation of~\cite{bena92}.
Alice uses weak coherent states, with phase encoding 0
or $\pi$ with respect to a strong coherent state. We denote
the weak states by $| \pm \alpha \rangle$ and the strong
state by $| \beta \rangle$. The overlap between the two states is
given by:
\begin{equation}
\label{overlap}
\left| \langle \alpha | -\alpha \rangle \right| \equiv \cos
\delta =
e^{-2|\alpha|^2} \; .
\end{equation}
Instead of sending the two states
one after the other, Alice uses two orthogonal polarizations:
$|\pm \alpha \rangle$, say, will have vertical polarization, and
$ |\beta \rangle$ will have horizontal polarization.
The decoding on Bob's side is schematized in
Fig.~\ref{detection}. The two
states are separated by a polarization beam splitter (PBS).
$ |\beta \rangle$ is rotated to vertical polarization and sent
through a mainly transmitting beam splitter (BS1) to detector
D1. A small fraction of $| \beta \rangle$, equal to
$| \alpha \rangle$, is sent to interfere with
$| \pm \alpha \rangle$
at BS2, and towards two detectors D2 and D3. A count in D2,
say, corresponds to phase zero, while a count in D3 corresponds
to $\pi$. No count in both D2 and D3 is of course an
inconclusive result. It is easy to calculate the probability
of such a result:
\begin{equation}
\label{prob?}
{\rm Prob}(?) = e^{-2|\alpha|^2} \; ,
\end{equation}
which is equal to the overlap between the two states. It was
shown that (\ref{prob?}) represents  the optimum for separating
deterministically two non-orthogonal
states~\cite{iva87}. Since all
the cases corresponding to inconclusive results have to be
discarded, the transmission rate of the channel is given
by:
\begin{equation}
\label{trans2}
t^{(2)} \equiv 1-{\rm Prob}(?) = 1 - e^{-2|\alpha|^2} \; .
\end{equation}
Detector D1 should always fire, and can therefore
be used as a trigger for the other two, enabling them only
for a short time corresponding to the length of the pulse
(this will reduce the dark counts). Moreover, an eavesdropper
with an inconclusive result would still have to send
$ | \beta \rangle$, which will result in random counts
in D2 and D3.
A small modification of the above will be used for our 4 + 2
protocol.

\subsection{Safety of the two-states systems}
\label{safety2}
As in Section~\ref{4}, we assume that Eve
performs an intercept/resend strategy, and  attempts to get
as much information as possible on the state sent by Alice.
This is obtained
by projecting it onto the orthogonal basis
$B_{\rm sym}= (i,j)$ as
shown in Fig.~\ref{projection}~\cite{per93}, and provides
Eve with probabilistic information only.
Another method for Eve would be to use the same POVM
as Bob. This would provide her with a smaller amount
of deterministic information.
The error rate introduced by Eve is equal to the probability
of making the wrong reading, e.g. the probability of obtaining
$j$ when the input was in fact $| \alpha \rangle$:
\begin{equation}
\label{ermax2}
q= {\rm Prob}(j/\alpha)=\frac{1-\sin \delta }{2} \; ,
\end{equation}
where $\delta$ is defined in (\ref{overlap}) and shown
in Fig.~\ref{projection}. With this particular
eavesdropping strategy, the transmission channel
Alice-Eve is known as a binary symmetric channel, and is
fully characterized by the error rate $q$~\cite{ash90}. The
mutual information shared by Alice and Eve is equal
to the channel capacity and is given by~\cite{eke94}:
\begin{equation}
\label{imax2}
i_{AE}(\delta) = 1+\frac{1-\sin \delta}{2}
\log_2 \left( \frac{1-\sin \delta}{2} \right) +
\frac{1+\sin \delta}{2}
\log_2 \left( \frac{1+\sin \delta}{2} \right) \; ,
\end{equation}
($i_{AE}(\delta)$ is the maximum
information that can be extracted from two states with overlap
$\cos \delta$). On the other hand, after Bob discards all
his inconclusive results, the transmission channel
Eve-Bob is perfect. Therefore the mutual information
Eve-Bob is:
\begin{equation}
i_{EB} = 1 \; .
\end{equation}
This strategy, where Eve eavesdrops on all the transmitted
photons, will create a high error rate.  In order to reduce it,
Eve shall only eavesdrop on a
fraction $\eta$ of the transmissions, thus creating an
error rate
\begin{equation}
\label{error2}
Q \equiv  \eta q =  \eta \frac{1-\sin \delta }{2}\; .
\end{equation}
It is now
easy to obtain the mutual information between Alice and Eve,
$I^{(2)}_{AE}$ and between Eve and Bob,
$I^{(2)}_{EB}$, as a function of the
error rate $Q$ and the angle $\delta$ between the states:
\begin{eqnarray}
I^{(2)}_{AE}(Q,\delta) &=&
\frac{2 Q}{1- \sin \delta} \; i_{AE}(\delta) \\
I^{(2)}_{EB}(Q,\delta) &=& \frac{2 Q}{1- \sin \delta}  \; ,
\end{eqnarray}
 Using
(\ref{overlap}) and (\ref{trans2}), we can also write the mutual
information as a function of the transmission rate for each
value of the error rate. These functions are plotted in
Fig~\ref{graph},
where the information gained for the 4-states system is used as
a reference.

\section{4 + 2 protocol}
\label{4+2}

\subsection{Principle and motivation}
The basic idea behind this protocol is that the 4-states
scheme does not require the two states in each basis to be
orthogonal. The safety of this scheme relies entirely on the
two points (i) and (ii) mentioned in Section~\ref{4}. Any
scheme using two pairs of states (each  pair corresponding
to one  non-orthonormal basis) satisfying both (i) and (ii) is an equally
good candidate for a 4-states protocol. Moreover, by choosing
non-orthogonal states in each pair, we will get the additional
advantage of the 2-states protocol, namely that Eve
cannot differentiate deterministically between the
two states in each basis.
As we show in the following, this will make the scheme
more sensitive to eavesdropping. A second advantage, shared with the
2-states system, is that this scheme will be more resilient
in the case of a lossy transmission line. This aspect will be
discussed in Section~\ref{lossy}.
A graphic comparison
between the different systems is given
in Fig~\ref{poincare}, where we plot the various states
on the Poincar\'{e} sphere.

\subsection{Implementation}
Let us apply this to our previous parallel reference phase encoding scheme.
The first pair  will
correspond to 0 and $\pi$ phase shifts, while the second
pair  will correspond to $\pi/2$
and $3 \pi/2$. Explicitly, the four states are:
$| \alpha\rangle$, $| -\alpha \rangle$,
$ |i \alpha \rangle$ and
$ |-i \alpha \rangle$. Let us emphasize that these states
satisfy condition (i) only when the intensity is low enough,
since pulses containing two or more photons do not
satisfy it. Indeed,
if we take into account the two photon number state, it is easy
to see that the two pairs: $ |\pm \alpha \rangle$ and
$| \pm i \alpha \rangle$ are not even in the same 2D Hilbert
space, and therefore cannot correspond to the same density
matrix. In the following, we shall assume that this condition
is verified. We shall however lift this restriction in
Section~\ref{lossy}, where we present an analysis for a lossy transmission
line.

The detection system is similar to the one explained in
Section~\ref{2}, with the addition of an optional $\pi/2$
phase shifter in one arm of the interferometer
(see Fig.~\ref{detection}).
When Bob wants to measure in the first basis, he does not
put the phase shift, and his detection scheme will therefore
differentiate between
$ |\alpha \rangle$ and $ |-\alpha \rangle$. When
he wants to measure in the second basis, he puts the $\pi/2$
phase shift, so that his detection scheme will now
differentiate between $|i \alpha \rangle$ and
$|-i \alpha \rangle$.
Of course, when Bob uses the wrong basis (say Alice sent
state $| \alpha \rangle$ and Bob puts the
$\pi/2$ phase shift),
Bob's result is totally uncorrelated with Alice's choice.
In many instances
Bob won't get any count in his detectors D2 and D3, which
correspond to inconclusive results, and will be discarded.

\subsection{Safety}
The problems facing Eve are now twofold:
she  doesn't know the basis used by Alice (as in the
4-states method), which means that even when she has a
conclusive result, she cannot be sure that it is relevant.
Moreover, as
the states are not orthogonal, she is faced with two
possibilities: either to try to get deterministic results,
which means that in many instances she will get no
information at all, but will still have to make a decision
on which state to send to Bob; or try to get probabilistic
information, in which case she will know very little on each
bit. In both cases, she will introduce errors, before she
even start to deal with the basis. This is the main difference
with the usual 4-states scheme: Alice and Bob make use of the fact
that they send weak pulses and not single photons
to enhance the safety.

In order to compare the various schemes,
we first need to calculate the transmission rate.
Since in half of the cases, Alice and Bob will use
different bases, and therefore will have to discard the transmission,
we easily get:
\begin{equation}
\label{trans42}
t^{(4+2)} = \frac{ 1-{\rm Prob}(?)}{2} =
\frac{1 - e^{-2|\alpha|^2}}{2}=\frac{1-\cos \delta}{2} \; .
\end{equation}
As in Section~\ref{4}, we now assume that Eve
eavesdrops on a fraction
$\eta$ of the transmissions, and uses the intercept/resend
strategy in the bases used by Alice and Bob.
When she uses the wrong basis, which happens on a fraction
$\eta/2$, she does not get any information, but still creates
an error rate $1/2$ as in the 4-states system. When she uses the
correct bases, again on a fraction $\eta/2$, the system
reduces to the 2-states system. Following Section~\ref{2},
we assume that Eve attempts to get as much
probabilistic information as possible. The error rate is thus
given by (\ref{ermax2}) and the information gain by
(\ref{imax2}). The overall error rate is therefore
\begin{equation}
\label{error42}
Q  =  \frac{\eta}{2} \left(1- \frac{\sin \delta }{2} \right) \; ,
\end{equation}
and the information gained by Eve
\begin{eqnarray}
I^{(4+2)}_{AE}(Q,\delta) &=&
\frac{ Q}{1- \frac{\sin \delta}{2}}
i_{AE}(\delta) \\
I^{(4+2)}_{EB}(Q,\delta) &=&
\frac{ Q}{1- \frac{\sin \delta}{2}}  \; .
\end{eqnarray}
The comparison between the three systems is given in
Fig~\ref{graph},
where we plot the various information as functions of the
transmission rate, and use the 4-states system as a reference.
Note that, as the transmission rate is different for the
2-states
system and the 4+2 system, we compare systems with different
values of $\alpha$ (or equivalently with different
values of the overlap
$\cos \delta$). With the type of eavesdropping strategy considered
here, the information available to Eve grows linearly with
the error rate for all the systems. Therefore, as the plot is
normalized with respect to the 4-states system, there is no
dependence on $Q$. Fig~\ref{graph} shows a clear
advantage for the 2-states system with respect to the
4-states system at very low transmission rates. However, the
information leaked to Eve increases rapidly with the
transmission rate. In contrast, the 4+2 system is always
preferable to both other systems. Let us emphasize that our
analysis is in principle correct for any value of the
overlap $\cos \delta$ between the states, if the various
states remain in a 2D Hilbert space. However, this is not
the case for the
suggested implementation with  weak coherent states. In this
case, the analysis is only appropriate for very weak states,
where the probability of having more than one photon is
sufficiently small, and therefore where the overlap between
the states is always close to one. This is likely to be
correct in practical applications, but it is still
important to present the corrections to the analysis when
two-photon states become relevant. This is the task of
Section\ref{lossy}.

\section{Lossy transmission line}
\label{lossy}
In the above work, we have analyzed the relative sensitivity
of various schemes to eavesdropping, under the assumption that
there was at most one photon per pulse. However, when we
use pulses of coherent or thermal light,
there is a non-zero probability to have
more than one photon in a pulse, even for
very weak pulses. When the transmission line between Alice and
Bob is approximately lossless, the fact that a few pulses
may have more than one photon is not too damaging, but only
gives some free information to an eavesdropper. However,
when the transmission line is lossy, using weak pulses can
prove disastrous for Alice and Bob. Let us assume that the
transmission line is a silica optical fiber, which is always
slightly lossy (about 0.2 dB/km).
If we want to use the cryptographic system over
reasonable distances, say up to 50 km, transmission losses
will be as high as 10 dB, or about 90\%. An eavesdropper
with superior
technology could replace the fiber by a perfectly transparent
one, and use the excess power for her mischievous purposes.

For the 4-states system based on polarization (Section~\ref{4}),
this can be very damaging.
Indeed, as polarization and photon number are independent
variables, there is no problem in principle to select the
few pulses with two or more photons and separate them into two
one-photon pulses, without changing the
polarization.
Eve could then send one of the two pulses to Bob, while keeping
the other one. She would measure her signal only after Alice
and Bob have disclosed their bases, which means that she
will get complete information on these pulses. If the pulses
sent by Alice are coherent and contain on average 1/10 photon, Eve would
obtain about one such pulse out of 200, and would send all
of these to Bob without attenuation. Remembering that Bob
only expect to receive one photon out of 100 pulses
sent by Alice,
we see that Eve can know the polarization of half of the
pulses received by Bob, without creating any error. For
thermal light, where the photon distribution is wider
(one pulse out of 100 would contain more than one photon),
Eve would actually know the polarization of
all the pulses. This can of course
be reduced by using even weaker pulses, but this in turn
would lower the transmission rate and the signal to noise
ratio (mainly because of the  dark counts of the detectors).
This shows that a combination of weak pulses with high
channel loss is a deadly combination for this cryptographic
system. If the losses are high enough, Eve can get full
information on the transmission, and this without causing
a single error. Let us emphasize however that any
system using only single photons would be insensitive to
the above attack.

In contrast to the 4-states system, both the 2-states systems
based on phase encoding (Section~\ref{2}),
and our new protocol (Section~\ref{4+2}) will
be more resilient against this attack. The main reason
is that as phase and number are conjugate variables, any
attempt by Eve to select the two-photon pulses will
unavoidably randomize the phase of the pulse. Therefore, the
best Eve can do is to split the incoming pulses, and send to
Bob the weaker pulses through a lossless fiber. For the
90 \% losses mentioned above, Eve can therefore split 90 \%
of the pulses. Bob of course has no way of discovering that
the losses are now due to Eve. The malicious Eve can now
keep her pulses, wait for Alice and Bob to disclose the basis
they used, and then try to extract the maximum information
>from them. In contrast to the 4-states system, her
information is limited by the fact that the states are not
orthogonal. In fact, the maximum information that she can
extract is given by $i_{AE}(\delta)$ defined in (\ref{imax2}).
For the above example, with 1/10 photon per pulse on average and
90 \% losses, Eve has pulses with 0.09 photons at her disposal.
The maximum information that she can extract is then about
0.23 bits. In a practical scheme, this would have to
be added to the information available to Eve as a function
of the error rate. Let us emphasize that Eve can adopt this beam
splitting strategy, even in the case of 4-states systems
with weak pulses. However, as was shown in the previous
paragraph, in this case she can also use the independence
of polarization and photon number to get more information.

\section{Conclusion}
\label{conclusion}
The rule of the game in quantum cryptography is that the
honest participants are limited by the available technology, while
the mischievous eavesdropper is only confined by the laws of
quantum  mechanics. This is in order to ensure that the safety
of the schemes does not depend on future technological advances,
no matter how unlikely. At present, one of the main
technological limitations is the difficulty of generating
single photons. For this reason, most existing schemes
rely on weak pulses, which are easy to generate. However,
in practical implementations~\cite{ben92,tow93,mul93},
this fact was considered to be
only a limitation. The fact that the pulses had a strong
vacuum component, or equivalently the fact that the two pulses
in the same basis were not orthogonal, was not utilized
by the users. In this work, we presented
a new system which made use of this vacuum component
to enhance the safety of the transmission.

The reason why the task of the eavesdropper is made more
difficult with this system, is that it combines the
strengths of the two systems it is built from: as in the
4-states system, Eve does not know the basis used by
Alice and Bob; as in the 2-states system, she cannot
distinguish with certainty between the two non-orthogonal
states. It is worth noting that the usual implementation
of the 4-states system with weak pulses actually
satisfies these two criteria. The weak pulses with
orthogonal polarizations are {\em not} orthogonal in the
Hilbert space, due to their common vacuum component.
The ingredient which is missing from these
system is the use of a strong reference pulse in the
detection, which ensures that Eve has to send a signal
to Bob, even when she could not get any information
on the signal sent by Alice. When this is added to the
system, the use of weak pulses becomes an advantage
for the legitimate users. Indeed, one
experimental realisation of quantum cryptography
based on an interferometric scheme~\cite{tow93},
is somewhat similar to our suggested implementation
(see Fig~\ref{detection}). The main difference between the two
schemes is that~\cite{tow93} do not use a strong
reference pulse (in their set-up, $|\beta|^2= |\alpha|^2$,
and the beam splitter BS1 is replaced by a mirror).
Therefore, as emphasized above, their scheme does
not take advantage of the non-orthogonality of the weak
pulses. Whenever she obtains an unsatisfactory result in
her attempt to eavesdrop, Eve may simply block the transmission
and send nothing to Bob. A slight modification of the
existing experiment could therefore provide a much safer
scheme.

Unfortunately, even very weak pulses have a non-zero
probability of having more than one photon. This creates
the most acute problem when the two legitimate users have
a lossy transmission line. In this case, we
showed that  the usual
polarization-based scheme appears to be unpractical, as too
much information may leak to an eavesdropper, without causing
any error. The two other schemes,
which combine non-orthogonal states, phase encoding and
a reference pulse,  appear to be less sensitive
to this problem, as Eve
cannot get total information, no matter what the losses are.
However, following the results of Section~\ref{lossy},
it is clear that, when the transmission line
between Alice and Bob is very lossy,
single photon transmission would be preferable.

\acknowledgments
We would like to thank  C.H. Bennett and A. Peres for
useful discussions.

\begin{figure}
\caption{Schematic of the detection system for 2-states
system and for 4+2 system. The description is given in the
text. The optional $\pi/2$ phase shift is used for the
4+2 system, and corresponds to the choice of decoding basis.}
\label{detection}
\end{figure}

\begin{figure}
\caption{Representation of two non-orthogonal states in the
Hilbert space. The maximum information that can be
extracted from them is obtained by projecting them
onto the basis $B_{\rm sym} = (i,j) $. A positive projection
on $i$ is read as input state $|\alpha \rangle$,
and reciprocally for $j$. However, as the projection of
$|\alpha \rangle$  on $j$ is non-zero, this creates errors
in the detection.}
\label{projection}
\end{figure}

\begin{figure}
\caption{Mutual information shared by Alice and Eve (full and dotted
curves) and by Eve and Bob (dashed and dash-dotted curves) as a function
of the transmission rate (in bit/pulse).
The dotted and dash-dotted curves refer to the
2-states system (Section~\protect\ref{2}); the full and dashed
ones to the 4+2 system (Section~\protect\ref{4+2}).
They are normalized with respect to
the mutual information for the 4-states system
(Section~\protect\ref{4}). This figure illustrates
the advantage of the new 4+2 protocol since the information
Eve may get is smaller, for any given transmission rate.
}
\label{graph}
\end{figure}

\begin{figure}
\caption{Difference between the different systems
shown by plotting the states on the Poincar\'{e} sphere.
The circles represent the 4-states system, where the two
states in basis $B_0$ (or $B_1$) are orthogonal, which
corresponds to  opposite directions on the Poincar\'{e} sphere.
The squares (or equivalently the triangles) represent
the 2-states system, where the states are not orthogonal,
and therefore cannot be distinguished deterministically.
The new 4+2 system makes use of both the triangles
(basis $B_0$) and the squares (basis $B_1$). }
\label{poincare}
\end{figure}

\end{document}